\documentclass[12pt]{article}

\usepackage{latexsym}

\font\tenmsbm=msbm10 scaled 1200
\font\sevenmsbm=msbm9
\newfam\msbmfam
\textfont\msbmfam=\tenmsbm
\scriptfont\msbmfam=\sevenmsbm

\makeatletter
\@addtoreset{equation}{section}
\makeatother

\def\beq{\begin{equation}}
\def\eeq{\end{equation}}
\def\bea{\begin{eqnarray}}
\def\eea{\end{eqnarray}}
\def\bet{\begin{tabular}}
\def\eet{\end{tabular}}

\catcode`@=11
\def\quad@rato#1#2{{\vcenter{\vbox{
        \hrule height#2pt
        \hbox{\vrule width#2pt height#1pt \kern#1pt \vrule width#2pt}
        \hrule height#2pt} }}}
\def\quadratello{\mathchoice
\quad@rato5{.5}\quad@rato5{.5}\quad@rato{3.5}{.35}\quad@rato{2.5}{.25} }

\begin{document}

\begin{titlepage}

\begin{flushright}

\end{flushright}

\vspace{1truecm}

\begin{center}

{\Large \bf  Classical Electromagnetic Field Theory in the Presence of
Magnetic Sources }

\vspace{2cm}
Wen-Jun Chen,$^a$
Kang Li$^{a}$\footnote{kangli@zimp.zju.edu.cn}
and Carlos M.
Na\'on $ ^{b}$\footnote{naon@venus.fisica.unlp.edu.ar}

\vspace{2cm}

{\it\small $^a$Zhejiang Institute of Modern Physics, Zhejiang
University, Hangzhou, 310027, P.R. China

\smallskip
$^b$Instituto de F\'{\i}sica La Plata, Departamento de F\'{\i}sica, Facultad de
Ciencias Exactas, Universidad Nacional de La Plata , CC 67,1900 La Plata, Argentina }

 \vspace{1cm}

\begin{abstract}
\vspace{0.5cm} Using two new well defined 4-dimensional potential vectors, we 
formulate the classical Maxwell's field theory in a form which has manifest Lorentz
covariance and $SO(2)$ duality symmetry in the presence of magnetic sources.  We set
up a consistent Lagrangian for the theory. Then from the action principle we get both
Maxwell's equation and the equation of motion of a dyon moving in the electro-magnetic
field.
\end{abstract}

\end{center}
\vskip 0.5truecm
\noindent

\end{titlepage}
\newpage
\baselineskip 6 mm

 \section{Introduction}

Recently there has been an increasing interest in the study of electromagnetic (EM)
duality symmetry, because it plays a fundamental role in superstring and brane theory
\cite{Kir} \cite{Schwarz-Seiberg}.  From Maxwell's equations we know that general EM
duality implies the existence of magnetic source ( magnetic charge (monopole) and
currents). However, when considering the quantum dynamics of particles carrying both
electric and magnetic charges (dyons) one faces the lack of a naturally defined
classical field theory despite of the fact that a consistent quantum field theory does
exist \cite{Brandt}. This issue was analyzed in recent contributions of many authors
\cite{Carneiro} \cite{LM} \cite{GM}\cite{Mendez}\cite{Galvao}. In our recent paper
\cite{LN1}, we gave an alternative formulation of electric-magnetic field theory in
the presence of magnetic source. The advantages of our formulation are the following.
First, we introduce two new potential vectors that have no singularities and we do not
need to use the concept of Dirac String; secondly, from the present paper we can set
up a consistent Lagrangian theory from which we can get all the information about
classical electromagnetic field theory which returns to the usual Maxwell's field
theory when only electric source is considered; thirdly it has manifest Lorentz
covariant and $SO(2)$ duality symmetry. Finally it seems that our formulation can be
quantized directly, an issue that will be reported in a forthcoming article.

The aim of this paper is to present the details of the construction of a Lagrangian
for the EM field theory in the formulation of \cite{LN1}. From the action principle we
expect to get the Maxwell's equation and also the equation of motion of a dyon moving
in the electro-magnetic field. We also explain also why our formalism has manifestly
$SO(2)$ duality symmetry. The paper is organized as follows. In the next section we
give a brief review of our formulation of classical electromagnetic field in the
presence of magnetic source, where two well defined 4-dimensional potential vectors
are introduced and the Maxwell's equations are written in a Lorentz covariant way. In
the third section we show how manifest $SO(2)$ duality symmetry arises in the present
approach. In section $4$, we give the Lagrangian form of our formulation. In section
5, from the action principle of the system of a dyon, we get both Maxwell's equation
and the equation of motion of the dyon. Some conclusion remarks are given in the last
section.

\section{Two potential vectors formulation}

Let us first give a brief review of the two 4-vector potentials formulation of the
electromagnetic field in the presence of magnetic source\cite{LN1}. Besides the usual
definition of 4-dimensional potential which we called $A_\mu^1$, i.e.
\begin{equation}
A_\mu^1=(\phi_1,-\mathbf{A_1}),~{\rm or}~ A^{\mu 1} =(\phi_1,\mathbf{A_1}),
\end{equation}
we also introduce
\begin{equation}
A_\mu^2=(\phi_2,-\mathbf{A_2}),~{\rm or}~ A^{\mu 2}=(\phi_2,\mathbf{A_2}),
\end{equation}
where $\phi_1$ and $\mathbf{A}_1$ are usual electric scalar potential and magnetic
vector potential in electrodynamics, while the newly introduced potential $\phi_2$  is
the scalar potential associated to  the magnetic field and $\mathbf{A}_2$ is a vector
potential associated to the electric field. It should be stressed that these two
4-potentials have no singularities around the magnetic charges (monopoles). Using
these potentials, the electric field strength $\mathbf{E}$ and the magnetic induction
$\mathbf{B}$ are then expressed as:

\begin{equation}
\mathbf{E}=-\nabla\phi_1 -\frac{\partial\mathbf{A_1}}{\partial
t}+\nabla\times\mathbf{A_2},
 \end{equation}
\begin{equation}
\mathbf{B}=\nabla\phi_2 +\frac{\partial\mathbf{A_2}}{\partial
t}+\nabla\times\mathbf{A_1}.
 \end{equation}
In the magnetic source free case, $\phi_2$ and $\mathbf{A}_2$ are
expected to be zero, so the above equation returns to the usual
magnetic source free case.

Now we introduce two field tensors as
\begin{equation}
F_{\mu\nu}^I=\partial_\mu A_\nu^I -\partial_\nu A_\mu^I,~~~~I=1,2.
\end{equation}
Then, choosing Lorentz gauge $\partial^\mu A_\mu^I=0 $, the Maxwell's equation in the
case of existing both electric and magnetic sources,
 \begin{equation}
 \nabla\cdot\mathbf{E}=\rho_e,~~~~~~\nabla\times\mathbf{B}=\mathbf{j_e}+\frac{\partial{\mathbf{E}}}{\partial{t}},
\end{equation}
\begin{equation}
\nabla\cdot\mathbf{B}=\rho_m,~~~~~~~~~~\nabla\times\mathbf{E}=-\mathbf{j_m}-\frac{\partial{\mathbf{B}}}{\partial{t}},
\end{equation}
can be recast as
\begin{equation}
\partial^\mu F_{\mu\nu}^I=g^{II'}J_\nu^{I'} ,
\end{equation}
where $$g^{II'}=\left(\begin{array}{cc}1&0\\0&-1\end{array}\right).$$ and
\begin{equation}
J_\mu^1=J_\mu^e=(\rho_e, -\mathbf{j_e}),~~J_\mu^2=J_\mu^m=(\rho_m, -\mathbf{j_m}).
\end{equation}
 In this formulation the currents are manifestly conserved:
\begin{equation}
\partial^\nu J_\nu^I \propto\partial^\nu\partial^\mu
F_{\mu\nu}^I=0.
\end{equation}

In the next section, we will see that the index $I$ is the $SO(2)$ index, so our
formulation above has manifestly $SO(2)$ duality symmetry which is related to the
general gauge transformation $A^{I}_{\mu} \rightarrow A^{I}_{\mu}+
\partial_{\mu}\chi^{I}$. The fields $\mathbf{E}$, $\mathbf{B}$, the field tensors in (2.5)
and Maxwell's equations (2.8) are all invariant under the transformations.

Let us stress also that in the expressions above neither $F_{\mu\nu}^1$ nor
$F_{\mu\nu}^2$ have the same matrix form  as the usual electromagnetic tensor. From
(2.5) and the
 definition (2.3) (2.4), we find that,
\begin{equation}
E_i=F_{0i}^1+{^\ast} F_{0i}^2
,
\end{equation}
and
\begin{equation}
B_i={^\ast}F_{0i}^1- F_{0i}^2
.
\end{equation}
So it is convenient to define a new field tensor as
\begin{equation}
F_{\mu\nu}=F_{\mu\nu}^1 +{^\ast} F_{\mu\nu}^2
,
\end{equation}
\begin{equation}
\widetilde{F}_{\mu\nu}={^\ast} F_{\mu\nu}^1 - F_{\mu\nu}^2 ,
\end{equation}
where $ \widetilde{F}_{\mu\nu}$ is exactly the Hodge star dual of $F_{\mu\nu}$. As we
shall see, using these new field tensors we can easily express the duality symmetry in
a compact fashion. It is easy to see that $F_{\mu\nu}$ is the analog to the usual
electromagnetic tensor defined in classical electrodynamics, because they have exactly
the same matrix form in terms of
 the field strengths. Since the vector
potentials in our formalism
 have no singularities one has $\partial^\mu
\,{^\ast}
 F_{\mu\nu}^I=0$, so Maxwell's equations can also be written as
\begin{equation}
\begin{array}{l}
\partial^\mu F_{\mu\nu}=\partial^\mu
F_{\mu\nu}^1=J_{\nu}^1\\
\partial^\mu \widetilde{F}_{\mu\nu}=-\partial^\mu
F_{\mu\nu}^2=J_{\nu}^2\\
\end{array}
\end{equation}
One can find that (refer to section 4), from $F_{\mu\nu}$ or $\widetilde{F}_{\mu\nu}$
defined above, we can easily build a Lagrangian such that the Maxwell's equation
(2.15) can be derived from the action principle.

\section{$SO(2)$ duality symmetry}

The $SO(2)$ duality symmetry of Electromagnetic field theory has been discussed in
many papers \cite{LM} \cite{Mendez} \cite{GM} \cite{Donev}. In our previous paper
\cite{LN1}, we explained in detail why the general duality symmetry is the $SO(2)$
symmetry, but there still exists something not very clear. For example, under the
general dual transformation for $F_{\mu\nu}, \widetilde{F}_{\mu\nu}$, i.e.
\begin{equation}
\left(\begin{array}{c}F_{\mu\nu}^{'}\\
\widetilde{F}_{\mu\nu}^{'}\end{array}\right)=\left(
\begin{array}{cc}a&b\\c&d\end{array}\right) \left(\begin{array}{c}F_{\mu\nu}\\
\widetilde{F}_{\mu\nu}\end{array}\right)
\end{equation}
why should the same transformation hold simultaneously for $J^{\mu 1} , J^{\mu 2}$?
One can make the same question concerning the dual transformations of
$(\mathbf{E},\mathbf{B})$ and $(q,g), (\mathbf{J}_e,\mathbf{J}_m)$, etc. Why all these
dual transformations must be the same? In our formulation, we can shed light on this
issue. So in this section we would first like to give answers to these questions and
then explain in detail why our formulation has manifestly $SO(2)$ duality symmetry,
i.e. we will see that the index $I$ of potentials $A_\mu^I$ is the $SO(2)$ index.

Let us first solve the Maxwell's equation in our formalism. It is easy to check that
the potential functions defined in the section above satisfy the differential
equations \cite{LN1}:
\begin{equation}
\begin{array}{l}
\frac{\partial^2}{\partial t^2} \phi_1 -\nabla^2\phi_1=\rho_e ,\\ ~\\
 \frac{\partial^2}{\partial
t^2}\mathbf{A_1}-\nabla^2\mathbf{A_1}=\mathbf{j_e}\\ ~~~~\\
\frac{\partial^2}{\partial
 t^2}\phi_2 -\nabla^2\phi_2=-\rho_m,\\
 ~~~~\\
\frac{\partial^2}{\partial
t^2}\mathbf{A_2}-\nabla^2\mathbf{A_2}=-\mathbf{j_m}\\
 \end{array}
\end{equation}

In the static case, i.e. when the sources do not depend on time $t$, we can write
\begin{equation}
\rho_I=\rho_I(\mathbf{x}),~~ \mathbf{J}_I=\mathbf{J}_I (\mathbf{x}),~~~~ I = 1 and 2,
 \end{equation}

where $I=1,2$ represent $I=e,m$ respectively.
 Then exactly as it is done in the standard classical
electrodynamics (magnetic source free case) \cite{Jackson}, the solution of equation
(3.2) is given by
\begin{equation}
\phi_I=\frac{1}{4\pi}g^{II'}\int \frac{\rho_{I'}(\mathbf{x}')}{r}d^3x'
\end{equation}
\begin{equation}
\mathbf{A}_I(\mathbf{x})=\frac{1}{4\pi}g^{II'}\int
\frac{\mathbf{J}_{I'}(\mathbf{x}')}{r}d^3x' ,\end{equation} where
$r=|\mathbf{x}-\mathbf{x'}|$, then from equations (2.3) and (2.4) we find that the
field strengths have the following representation

\begin{equation}
\mathbf{E}(\mathbf{x}) =\frac{1}{4\pi}\int\rho_e
(\mathbf{x}')\frac{\mathbf{r}}{r^3}\,d^3x'\\ +\frac{1}{4\pi}\int\mathbf{J}_m
(\mathbf{x}')\,{\bf \times}\,\frac{\mathbf{r}}{r^3}\,d^3x
\end{equation}
\noindent and
\begin{equation}
\mathbf{B}(\mathbf{x}) =\frac{1}{4\pi}\int\rho_m
(\mathbf{x}')\frac{\mathbf{r}}{r^3}\,d^3x'\\ -\frac{1}{4\pi}\int\mathbf{J}_e
(\mathbf{x}')\,{\bf \times}\,\frac{\mathbf{r}}{r^3}\,d^3x.
\end{equation}

Now we can give the answer to the question in the beginning of this section. Because
$E_i=F_{0i}, B_i = \widetilde{F}_{0i}$, then we know that if $F_{\mu\nu},
\widetilde{F}_{\mu\nu}$ have a transformation given by equation (3.1), which led to
the field strengths $\mathbf{E}, \mathbf{B}$ having the same transformation, and
because the field strengths are related to the sources by equation (3.6) and (3.7), so
the sources $\rho_e,\rho_m$ and $\mathbf{J}_e,\mathbf{J}_m$ must change in the same
way. The same transformation must be satisfied by the 4-dimensional currents $J^{\mu
1},J^{\mu 2}$. That is why once one chooses the transformation for $F_{\mu\nu},
\widetilde{F}_{\mu\nu}$ given by (3.1), then the corresponding field strengths and the
the sources must obey the same transformation. If we impose that the Maxwell's
equations (2.6) and (2.7) are invariant under these transformations of field strengths
and the sources, we obtain
 $a = d$ and $b = -c$. Moreover, if we also impose that the energy
density $\frac{1}{2}(\mathbf{E}^2+\mathbf{B}^2)$ and the Poynting vector
$\mathbf{E}\times\mathbf{B} $ are invariant under this transformation we get $a^2 +
b^2 =1$. It is then natural to introduce an angle $\alpha$ such that $a=\cos\alpha$
and $b=\sin\alpha$. Hence the general duality transformation matrix coincides with the
general rotation matrix in two dimensions. Thus it becomes apparent that the general
electro-magnetic duality symmetry is the SO(2) symmetry. Under the special case,
$\alpha =\pi /2$, the transformation (3.1) coincides with the replacement
$F_{\mu\nu}\rightarrow \widetilde{F}_{\mu\nu},~~\widetilde{F}_{\mu\nu}\rightarrow
-F_{\mu\nu}$ and the same replacements must be taken simultaneously, i.e.
$\mathbf{E}\rightarrow\mathbf{B},\mathbf{B}\rightarrow
-\mathbf{E},~\rho_e\rightarrow\rho_m , \rho_m\rightarrow -\rho_e$ and
$\mathbf{J}_e\rightarrow\mathbf{J}_m, \mathbf{J}_m\rightarrow -\mathbf{J}_e$, etc.
This corresponds to the  usual special electro-magnetic duality symmetry.

Now we would like to point out that the index $I$ of the potentials $A_\mu^I$ is the
$SO(2)$ index. Under the general dual transformation, i.e. $SO(2)$ transformation,
from the discussion above we know that the sources $\rho_I$ and $\mathbf{J}_I$ change
as
\begin{eqnarray}
\rho_{I'}=R(\alpha)_{I I'}\rho^{I'}\\
\mathbf{J'}_I=R(\alpha)_{I I'}\mathbf{J}^{I'},
\end{eqnarray}
where $R(\alpha)$ is the $SO(2)$ rotation matrix. Then from equations (3.4) and (3.5),
we know that the potentials $A_{\mu}^I=(\phi_I, -\mathbf{A}_I)$ should have the same
$SO(2)$ transformation, that is to say that the index $I$ of potential $A_\mu^I$ is
the $SO(2)$ index. So our formulation in section (2) has manifestly $SO(2)$ duality
symmetry.

\section{Lagrangian formulation of the field}
In this section we will give a lagrangian for the electromagnetic field which gives
the right Maxwell's equation in the presence of magnetic source. We will also see that
from this lagrangian one can deduce the right Lorentz force formula (refer to the next
section).

 The Lagrangian of the field is given by
\begin{equation}
 L=-\frac{1}{4}(F_{\mu\nu})^2-(A_\mu^1+\ast
A_\mu^2)J^{\mu 1}+(\ast
 A_{\mu}^1-A_\mu^2)J^{\mu 2},
 \end{equation}
\noindent where the $\ast A_{\mu}^I$ is defined through
\begin{equation}
\partial_\mu\ast
A_\nu^I=\frac{1}{2}\epsilon_{\mu\nu}^{~~\alpha\beta}\partial_\alpha A_\beta^I.
\end{equation}

From a simple calculation we find
\begin{equation}
\frac{\partial F^2}{\partial (\partial_\mu A_\nu^1)}
   =\frac{\partial
(F_{\alpha\beta}F^{\alpha\beta})}{\partial (\partial_\mu A_\nu^1)}=4 F^{\mu\nu},
 \end{equation}

\begin{equation}
\frac{\partial F^2}{\partial (\partial_\mu A_\nu^2)}
   =\frac{\partial
(F_{\alpha\beta}F^{\alpha\beta})}{\partial (\partial_\mu A_\nu^2)}=4
\widetilde{F}^{\mu\nu}.   \end{equation}

Notice that $\ast A_{\mu}^I$ is related to the derivative of $A_{\alpha}^I $, and also
take into account the conservation conditions of the currents (2.10). Then the
Euler-Lagrange equation of the Lagrangian defined in (4.1) gives the Maxwell's
equation
 (2.15).

\section{Action principle of the field}

 In this section, we would like to give a consistent action
 principle formulation for the classical electro-magnetic field
 theory in the presence of the magnetic sources. We can consider a
 system of dyons which interact with the electro-magnetic field.
 From the action of the system we expect to get both field
 equation (2.15) and also the equations of motion of the dyons.

 For simplicity, we consider here one dyon with electric
 charge $q$ and magnetic charge $g$, which moves in the electro-magnetic field (the
 extension to the many dyons system can be easily done). The action of this system
   consists of three parts, i.e,
\begin{equation}
  S=S_p +S_I+S_F,
  \end{equation}
\noindent  where,
  \begin{equation}
  S_p =-m\int_{\lambda_1}^{\lambda_2}\sqrt{-g_{\mu\nu}\frac{dx^\mu}{d\lambda}
   \frac{dx^\nu}{d\lambda}}d\lambda
  \end{equation}
\noindent is the free action of the dyon, and
\begin{eqnarray}
S_I=q\int_{\lambda_1}^{\lambda_2}(A_\mu^1+\ast A_\mu^2)\frac{dx^\mu}{d\lambda}d\lambda
-g\int_{\lambda_1}^{\lambda_2}(\ast A_\mu^1- A_\mu^2)\frac{dx^\mu}{d\lambda}d\lambda
\nonumber&~&\\ =\int_\Omega (A_\mu^1+\ast A_\mu^2)J^{\mu 1}d^4x-\int_\Omega (\ast
A_\mu^1- A_\mu^2)J^{\mu 2}d^4x&~&
\end{eqnarray}
\noindent is the term of interaction between the dyon and the electro-magnetic field
around it. The $J^{\mu I}$ in the above equation are the currents for one particle
dyon which have the form
\begin{eqnarray}
J^{\mu 1}=q\int \frac{dx^\mu}{d\tau}\delta^4(x-x(\tau))d\tau ,&`&\\ J^{\mu 2}=g\int
\frac{dx^\mu}{d\tau}\delta^4(x-x(\tau))d\tau.
\end{eqnarray}
The last term of the action
\begin{equation}
S_F=-\frac{1}{4}\int_\Omega F_{\mu\nu}F^{\mu\nu}d^4x
\end{equation}
\noindent is nothing but the action of the electro-magnetic field.

Let us now vary the potentials as
\begin{equation}
A_\mu^I\Rightarrow A_\mu^I+\epsilon_I B_\mu^I(x),~~ B_\mu^I|_\Omega =0.
\end{equation}

We can check that
\begin{equation}
\frac{\partial
S_f}{\partial\epsilon_1}|_{\epsilon_1=\epsilon_2=0}=\int_\Omega\partial_\mu
F^{\mu\nu}B_\nu^1d^4x
\end{equation}
\begin{equation}
\frac{\partial
S_f}{\partial\epsilon_2}|_{\epsilon_1=\epsilon_2=0}=\int_\Omega\partial_\mu
^{\ast}F^{\mu\nu}B_\nu^2d^4x
\end{equation}
Noticing that $\frac{\partial \ast A_\mu^I}{\partial \epsilon^{I'}}=0$ and
$\frac{\partial S_p}{\partial\epsilon_I}=0$, we then have
\begin{equation}
0=\frac{\partial S}{\partial\epsilon_1}|_{\epsilon_1=\epsilon_2=0}=\int_\Omega
(\partial_\mu F^{\mu\nu}-J^{\nu 1})B_\mu^1d^4x,
\end{equation}
\begin{equation}
0=\frac{\partial S}{\partial\epsilon_2}|_{\epsilon_1=\epsilon_2=0}=\int_\Omega
(\partial_\mu \widetilde{F}^{\mu\nu}-J^{\nu 2})B_\mu^1d^4x.
\end{equation}
 Because $B_\mu^1$ and $B_\mu^2$ are arbitrary, then from the equations
 (5.10) and (5.11) above we reobtain the Maxwell's equation (2.15).

 Further if we change the coordinate of the dyon in the form
 \begin{equation}
 x^\mu\Rightarrow x^\mu +\epsilon_0 y^\mu, {\rm
 with}~y^\mu(\lambda_1)=y^\mu (\lambda_2)=0,
 \end{equation}
\noindent we find
\begin{equation}
\frac{\partial
S}{\partial\epsilon_0}|_{\epsilon_0=0}=\int_{\tau_1}^{\tau_2}(-m\frac{d^2x^\alpha}{d\tau^2}
+q F^{\alpha}_{~\beta}\frac{dx^\beta}{d\tau}- g
\widetilde{F}^{\alpha}_{~\beta}\frac{dx^\beta}{d\tau})y_\alpha . d\tau
\end{equation}
Since $y_\alpha$ is arbitrary, then from $\frac{\partial
S}{\partial\epsilon_0}|_{\epsilon_0=0}=0$ we obtain
\begin{equation}
m\frac{d^2x^\alpha}{d\tau^2}= q F^{\alpha}_{~\beta}\frac{dx^\beta}{d\tau}- g
\widetilde{F}^{\alpha}_{~\beta}\frac{dx^\beta}{d\tau}.
\end{equation}
This is just the equation of motion of a dyon moving in the electro-magnetic field.
From this equation we can find that the Lorentz force the dyon acquired in the
magnetic field can be represented in terms of field strengths as
\begin{equation}
\mathbf{F}=q\,(\mathbf{E}+\mathbf{v}\times\mathbf{B})+g\,(\mathbf{B}-\mathbf{v}\times\mathbf{E}).
\end{equation}

We would like to stress that the general Lorentz force has also the $SO(2)$
electro-magnetic duality symmetry.

\section{Conclusion remarks}

The main results of this paper are as follows. First we use the formulation of
reference \cite{LN1} to explain why the classical electro-magnetic field theory in the
presence of a magnetic source has exactly the $SO(2)$ duality symmetry. Then we find a
proper Lagrangian formulation for the theory, and at last using the action principle
of the system of dyons we derived both Maxwell's equation and the equation of motion
for the dyon. From this equation of motion we got the general Lorentz force for a dyon
moving in the electro-magnetic field.

As a consistency check of our formulation we see that for $g=0$ and $J^{\mu 2}=0$ (no
magnetic sources), from equation (3.2) we can set $A_\mu^2=0$, and so
$F_{\mu\nu}=F_{\mu\nu}^1+^{\ast}F_{\mu\nu}^2\Rightarrow F_{\mu\nu}^1$. This means that
our formulation contains standard electrodynamics as a particular case. For $q=0$ and
$J^{\mu 1}=0$ (no electric sources), one has $A_\mu^1=0$, and then
$F_{\mu\nu}=F_{\mu\nu}^1+^{\ast}F_{\mu\nu}^2\Rightarrow ^{\ast}F_{\mu\nu}^2$, and the
lagrangian becomes:
\begin{equation}
L=-\frac{1}{4}(^\ast F_{\mu\nu}^2)^2-A_{\mu}^2 J^{\mu
2}=\frac{1}{4}(F_{\mu\nu}^2)^2-A_{\mu}^2 J^{\mu 2}.
\end{equation}
Thus in this case the formulation is completely parallel to the magnetic source free
case.

\paragraph{Acknowledgements.}

We would like to thank S. Carneiro for his kind and valuable comments and for letting
us know reference \cite{Carneiro}. Part of this work is done during K. Li 's research
visiting to Abdus Salam International Centre for Theoretical Physics, Trieste Italy.
This work was partially supported by the National Nature Science Foundation of China
and the Consejo Nacional de Investigaciones Cient\'{\i}ficas y T\'ecnicas (CONICET),
Argentina.

\end{document}